\documentclass[conference]{IEEEtran}
\IEEEoverridecommandlockouts
\usepackage{cite}
\usepackage{amsmath,amssymb,amsfonts}
\usepackage{algorithmic}
\usepackage{graphicx}
\usepackage{textcomp}
\usepackage{xcolor}
\usepackage{subfloat}
\usepackage{subfig}
\usepackage{makecell,rotating}
\usepackage{multirow}
\usepackage{booktabs}
\usepackage{hyperref}
\usepackage{cuted}\stripsep-3pt
\usepackage[misc]{ifsym} 
\def\BibTeX{{\rm B\kern-.05em{\sc i\kern-.025em b}\kern-.08em
    T\kern-.1667em\lower.7ex\hbox{E}\kern-.125emX}}
\begin{document}

\title{CM-MLP: Cascade Multi-scale MLP with Axial Context Relation Encoder for Edge Segmentation of Medical Image\\
}
\renewcommand{\thefootnote}{\fnsymbol{footnote}}

\DeclareRobustCommand*{\IEEEauthorrefmark}[1]{%
	\raisebox{0pt}[0pt][0pt]{\textsuperscript{\footnotesize\ensuremath{#1}}}}
\author{\IEEEauthorblockN{Jinkai Lv\IEEEauthorrefmark{1}\footnotemark{$^\dagger$}, Yuyong Hu\IEEEauthorrefmark{1}\footnotemark{$^\dagger$}, Quanshui Fu\IEEEauthorrefmark{3}\footnotemark{$^\dagger$}, Zhiwang Zhang\IEEEauthorrefmark{5}, Yuqiang Hu\IEEEauthorrefmark{6}, Lin Lv\IEEEauthorrefmark{7} \\Guoqing Yang\IEEEauthorrefmark{3}\footnotemark{*}, Jinpeng Li\IEEEauthorrefmark{4}\footnotemark{*}, and Yi Zhao\IEEEauthorrefmark{1,2}\footnotemark{*}}
	
	\IEEEauthorblockA{\IEEEauthorrefmark{1}Henan Institute of Advanced Technology,
		Zhengzhou University, Zhengzhou, China}
	\IEEEauthorblockA{\IEEEauthorrefmark{2}The Research Center for Ubiquitous Computing Systems, Institute of Computing Technology, \\Chinese Academy of Sciences, Beijing, China}
	\IEEEauthorblockA{\IEEEauthorrefmark{3}Suining Central Hospital, Suining, China}
	\IEEEauthorblockA{\IEEEauthorrefmark{4}Hwa Mei Hospital, University of Chinese Academy of Sciences, Ningbo ,China}
	\IEEEauthorblockA{\IEEEauthorrefmark{5}School of Electrical and Information Engineering, The University of Sydney, Sydney, Australia}
	\IEEEauthorblockA{\IEEEauthorrefmark{6}School of Mathematics and Computer Sciences, NanChang University, Nanchang, China}
	\IEEEauthorblockA{\IEEEauthorrefmark{7}School of Information and Software Engineering, University of Electronic Science and Technology of China, Chengdu, China}
	\IEEEauthorblockA{snszxyy@126.com, lijinpeng@ucas.ac.cn, biozy@ict.ac.cn}
}

\maketitle

\footnotetext[2]{These authors contributed equally to this work. }
\footnotetext[1]{Corresponding author. }

\begin{abstract}
	
The convolutional-based methods provide good segmentation performance in the medical image segmentation task. However, those methods have the following challenges when dealing with the edges of the medical images: (1) Previous convolutional-based methods do not focus on the boundary relationship between foreground and background around the segmentation edge, which leads to the degradation of segmentation performance when the edge changes complexly. (2) The inductive bias of the convolutional layer cannot be adapted to complex edge changes and the aggregation of multiple-segmented areas, resulting in its performance improvement mostly limited to segmenting the body of segmented areas instead of the edge. To address these challenges, we propose the CM-MLP framework on MFI (Multi-scale Feature Interaction) block and ACRE (axial context relation encoder) block for accurate segmentation of the edge of medical image. In the MFI block, we propose the cascade multi-scale MLP (Cascade MLP) to process all local information from the deeper layers of the network simultaneously and utilize a cascade multi-scale mechanism to fuse discrete local information gradually. Then, the ACRE block is used to make the deep supervision focus on exploring the boundary relationship between foreground and background to modify the edge of the medical image. The segmentation accuracy (Dice) of our proposed CM-MLP framework reaches 96.96\%, 96.76\%, and 82.54\% on three benchmark datasets: CVC-ClinicDB dataset, sub-Kvasir dataset, and our in-house dataset, respectively, which significantly outperform the state-of-the-art method. The source code and trained models will be available at https://github.com/ProgrammerHyy/CM-MLP.

\end{abstract}


\section{INTRODUCTION}
In clinical diagnosis, medical image segmentation is a primary task, which has been extensively studied by the medical imaging community\cite{b1,b2,b3}. Compared with the traditional manual labelling method, the medical image segmentation algorithm\cite{b4,b5} can help doctors quickly find the location of lesions and reduce the workload. Therefore, the medical image segmentation algorithm is vital in medical image processing and analysis.

In recent years, more and more researchers have applied convolutional layers to medical image segmentation tasks\cite{b4,b5,b6,b7}. Oktay \textit{et al.} proposed a landmark work called UNet\cite{b8}. UNet\cite{b8} contains a U-shaped encoder-decoder architecture using a pyramid-like sampling process and skip connections to preserve the low-level semantic information. Following UNet, many different convolutional neural networks have been proposed, such as UNet++\cite{b4}, UNet3+\cite{b9}, 3D UNet\cite{b10}, V-Net\cite{b11}, Y-Net\cite{b12}, and KiUNet\cite{b13}. However, these popular Unet-based structures focus on improving the overall segmentation performance rather than extracting the edge of medical image information, which is crucial for improving the segmentation performance of the network.

Recently, MLP-based (Multilayer Perceptron based) methods\cite{b14,b15,b16,b17,b18,b19,b20} provide promising results in computer vision tasks. MLP-Mixer\cite{b14} demonstrated that without convolution layers and self-attention mechanisms, it can still provide comparable performance to less computationally intensive transformers-based methods\cite{b21,b22}. Maxim\cite{b19} further applied MLPs to low-level vision tasks and achieved satisfactory performance. MLP-based methods\cite{b14,b15,b16,b17,b18,b19,b20} overcome the inductive bias of the weights and can process all the local information of the image at same time. Therefore, these MLP-based methods can naturally solve the problem of insufficient edge information extraction by the recent popular network.

Inspired by these MLP-based methods\cite{b14,b15,b16,b17,b18,b19,b20}, we propose Cascade Multi-scale MLP (CM-MLP) framework, which is a denser design architecture with Multi-scale Feature Interaction block (MFI) and axial context relation encoder (ACRE). Using the MFI block, CM-MLP can overcome the influence of inductive bias brought by convolution layers, simultaneously process all local information and gradually fuse the discrete local information. Using the ACRE block, CM-MLP can focus on the boundary relationship between foreground and background around the segmentation edge.

 The contributions of this paper can be summarized as follows:

\begin{itemize}
	\item We propose a novel CM-MLP framework to extract better edge information in medical image segmentation. In this framework, we proposed the MFI block to capture complex and dense edge (i.e., aggregated multiple segmented regions) information. MFI block can process all local information simultaneously and gradually fuse the discrete local information. In addition, we also proposed the ACRE block to make the CM-MLP framework focus on segmenting object edges rather than bodies. With the cooperation of the MFI block and ACRE block, CM-MLP overcomes the influence of inductive bias brought by convolution layers and neglect of the boundary relationship between foreground and background.

	\item Comparison results show that our proposed CM-MLP framework outperforms the previous state-of-the-art method on the CVC-ClinicDB dataset, sub-Kvasir dataset, and our in-house dataset.

\end{itemize}

\section{RELATED WORK}

\subsection{Convolutional model}

The emergence of medical image segmentation frameworks based on CNN (Convolutional Neural Network) models, especially Unet\cite{b8}, pioneered segmentation networks with convolution as the main architecture. DUNet\cite{b23}, based on the U-Net framework, uses a Deformable Convolution Block\cite{b24} as each unit of the encoder and decoder. The deformable convolution block simulates different shapes and scales by learning local, dense, and adaptive receptive fields. R2U-Net\cite{b25} combines residual connections and recurrent convolutions to replace the original submodules in U-Net. BIO-Net\cite{b26} proposed bi-directional skip connections to extract more spatial information by recurrently reusing the building blocks and then using the architecture optimization algorithm in BIX-NAS\cite{b27} to optimize the connection in a multi-stage network. PraNet\cite{b28} establish the relationship between areas and boundary cues using the reverse attention (RA) module, and then CaraNet\cite{b29} use Channel-wise Feature Pyramid (CFP) module with A-RA(Axial reverse attention) to further mine the boundary cues. However, in the reverse attention, only the pixels around the segmentation result are highlighted. The error pixels may be kept in the final result because the pixels have been wrong segmented in the previous operation while not fixing the error pixels.

With the continuous development of the convolutional network, the performance of which is increasingly affected by the complex edge information. However, all of the above attempts\cite{b26,b27,b28,b29} are still based on convolutional-based architectures. It is inductive bias and neglect of the boundary relationship between foreground and background around the segmentation edge hindering the network's ability to extract edge information. Therefore, we take advantage of the ability of MLP to process all local information simultaneously and focus on the boundary relationship between foreground and background around the segmentation edge, thereby improving the network's ability to extract complex edge information.

\subsection{MLP-based model}

MLP-Mixer\cite{b14}, an MLP-based architecture that replaces self-attention with simple token-mixing MLP, achieves competitive results in image classification. gMLP\cite{b15} demonstrated that self-attention is not necessary for NLP tasks through gated-based MLP. CYCLEMLP\cite{b16} achieves a linear computational complexity related to image size through CycleFC, enabling pure MLP architectures for object detection and segmentation in larger images. MLP-3D\cite{b17} leverages VISION PERMUTATOR\cite{b18} in the video classification task to encode feature representations with linear projections along the height and width dimensions, respectively. Then, giving token-mixing MLP a temporal modelling Ability by GTM (Grouped Time Mixing) in the temporal dimension. MAXIM\cite{b19}, as a network based on a pure MLP architecture, utilizes a cross-gated module and a multi-axis gated MLP to achieve the mixing of local and global spatial information. RepMLPNet\cite{b20} merges the trained parameters of the parallel convolution kernels into the FC kernel and merges the local prior into the FC (Full Connect) layer utilizing local injection. Therefore, the RepMLPNet can capture local and global information and becomes a pure MLP-structured model in the inference stage.

When using MLP for vision tasks, we have noticed that some appropriate designs can make MLP even more potent than convolutions. Therefore, we propose the CM-MLP framework based on the MFI (Multi-scale Feature Interaction) block and ACRE (axial context relation encoder) block for accurate segmentation of the edge of the medical image. The MFI block can process all local information simultaneously by Cascade MLP (cascade Multi-scale MLP). Then the ACRE block can help our CM-MLP framework focus on the boundary relationship between foreground and background around the segmentation edge to better extract the edge information. 

Note that although the computational complexity of the MFI block is linearly related to the H$\times$W same as \cite{b19}, considering the size of the feature map and the amount of information contained, we only apply the MFI block in the last three layers of the network to further reduce the amount of parameters. The computational cost is negligible.

\begin{figure*}[!t]
	\centering
	\subfloat{\includegraphics[scale=0.6]{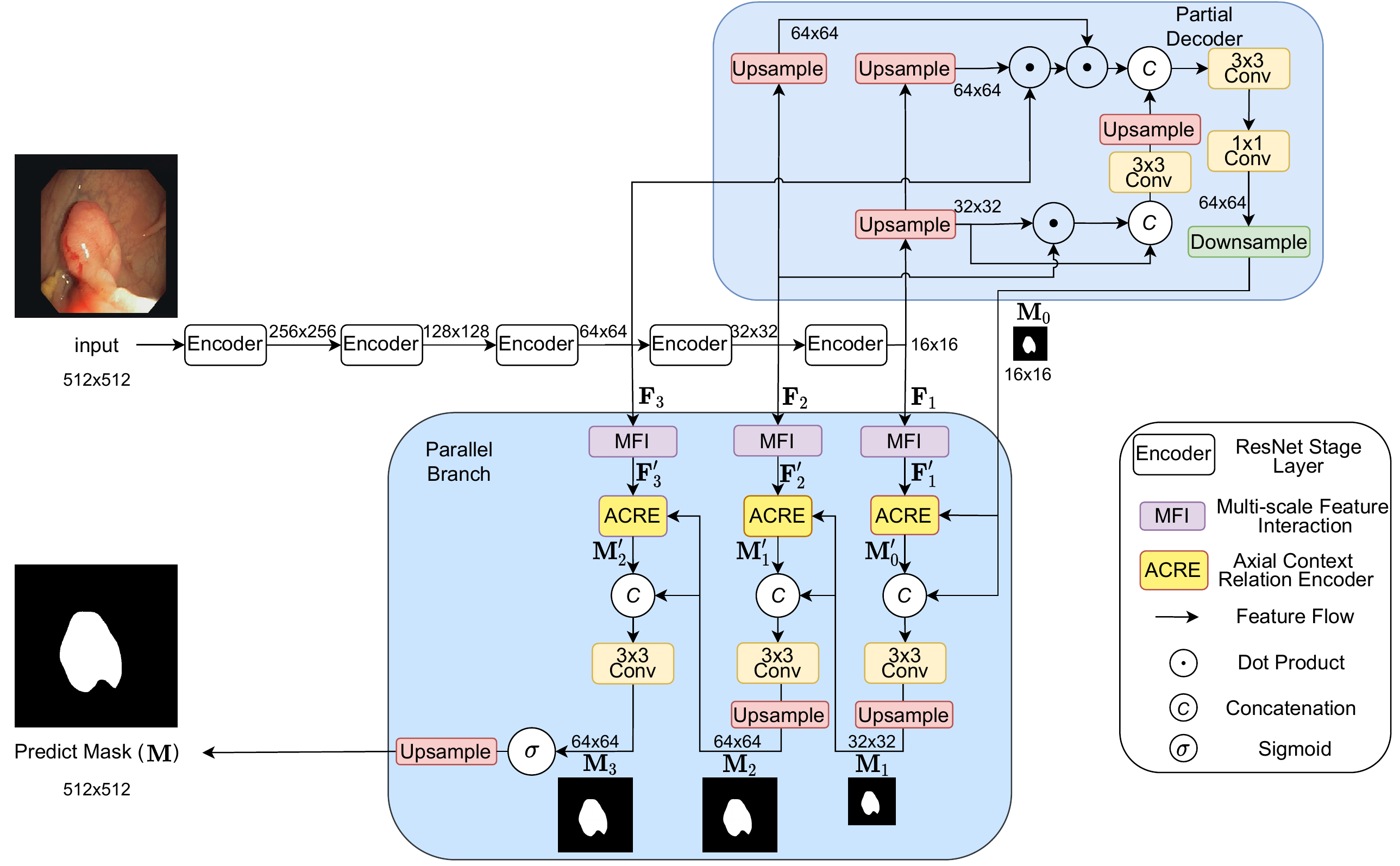}}
	\caption{Overview of the proposed CM-MLP framework, which consists of the 5-stage Encoder, the Partial Decoder, and the Parallel Branch. The 5-stage Encoder first encodes the input image. The feature maps from the last three stages (i.e., $\mathbf{F}_1, \mathbf{F}_2, \mathbf{F}_3$) are decoded by the Partial Decoder to produce the original mask $\textbf{M}_0$. In the Partial Decoder, the feature maps will dot products of each other in the same size by upsampling and generate the original mask $\textbf{M}_0$ by concatenated operation, convolution layer, and downsampling operation. In the first branch of the Parallel Branch, the MFI block takes feature map $\mathbf{F}_1$ as the input to get feature map $\mathbf{F}'_1$. Then the ACRE block takes $\mathbf{F}'_1$ and $\textbf{M}_0$ as input and generates the refined mask $\textbf{M}'_0$. The higher resolution mask $\textbf{M}_1$ is generated from refined mask $\textbf{M}'_0$ and original mask $\textbf{M}_0$ by concatenated operation, convolution layer, and upsampling operation. The other two branches go through the same operation. The final mask $\textbf{M}$ is obtained by $\textbf{M}_3$ using sigmoid and upsampling operations. In Deep Supervision, after having $\textbf{M}_0$, $\textbf{M}_1$, $\textbf{M}_2$ and $\textbf{M}_3$, we upsample those results to the same size as ground truth and calculate the total loss.}
	\label{fig_1}
\end{figure*}

\section{METHOD}\label{method}

We propose the CM-MLP framework for medical image segmentation and adopt the MFI and ACRE blocks. Unlike\cite{b30}, which added MLP to the encoder and decoder of Unet, MLPs are in parallel with Unet in our proposed CM-MLP framework. In \ref{AA}, we briefly introduce the CM-MLP framework. In \ref{AB}, we introduce the principle of the MFI block, which can process all local information simultaneously and gradually fuse discrete local information. In \ref{AC}, we introduce the principle of the ACRE block, which can focus on the boundary relationship between foreground and background around the segmentation edge. In \ref{AD}, we introduce the loss function of our CM-MLP framework.

\begin{figure*}[!t]
	\centering
	\subfloat{\includegraphics[scale=0.5]{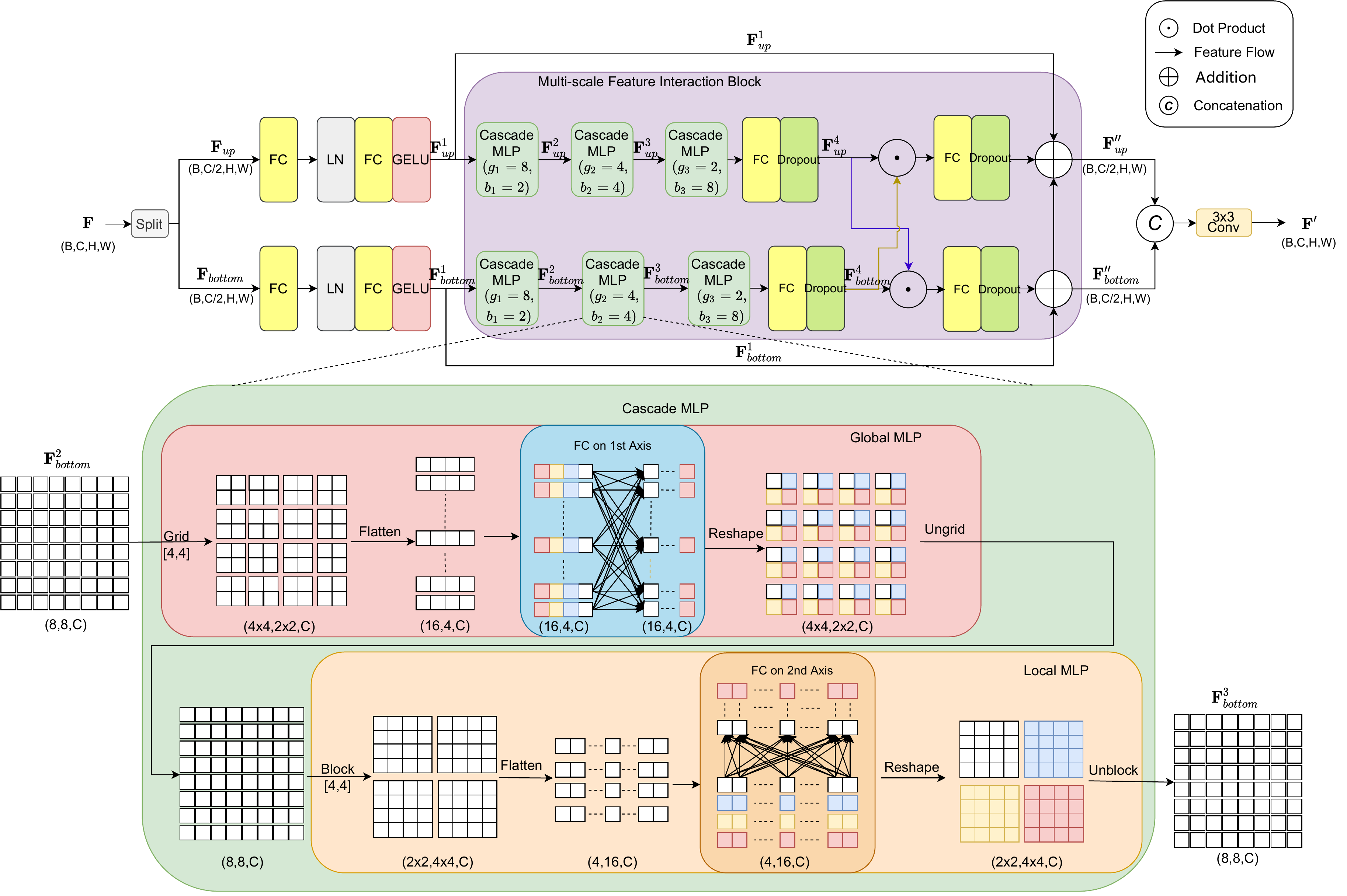}}
\caption{Illustration of the Multi-scale Feature Interaction (MFI) block. The input feature map $\mathbf{F}$ is channel-wise split into two branches $\mathbf{F}_{up}$ and $\mathbf{F}_{bottom}$. After each branch processed by the multiple Cascade MLP blocks, it will be alternately multiplied to increase information interaction and added together ($\mathbf{F}''_{up} = \mathbf{F}^1_{up} + \mathbf{F}^4_{up}\times \mathbf{F}^4_{bottom} + \mathbf{F}''_{bottom}, \mathbf{F}''_{bottom} = \mathbf{F}^1_{bottom} + \mathbf{F}^4_{bottom}\times \mathbf{F}^4_{up}$). The output of MFI block $\mathbf{F}'$ is obtained by concatenating two branch features $\mathbf{F}''_{up}$ and $\mathbf{F}''_{bottom}$ and the convolution layer. For Cascade MLP, we take the second Cascade MLP in MFI block ($b_2 = 4$, $g_2 = 4$) as an example. For better understanding, we used $\mathbf{F}^2_{bottom}$ $(W=8, H=8, C)$ as input, where C is the size of the channel. Input feature $\mathbf{F}^2_{bottom}$ will be processed by Global MLP and Local MLP to obtain the output feature $\mathbf{F}^3_{bottom}$. In Global MLP, the feature map is first grid into $4\times 4$ ($g_2 = 4$) non-overlapping patches which sizes is $2\times 2$. After flattening, the FC layer is executed on the first axis (the same colour represents the FC layer's input and output vector) and then reshaped back and Ungrid to the original size. In Local MLP, the feature map is first blocked into $2\times 2$ non-overlapping patches which sizes is $4\times 4$ ($b_2 = 4$). After flattening, the FC layer is executed on the second axis and then reshaped back and unblock to the original size to get $\mathbf{F}^3_{bottom}$ feature map.}
	\label{fig_2}
\end{figure*}

\begin{figure}[!t]
	\centering
	\subfloat{\includegraphics[scale=0.48]{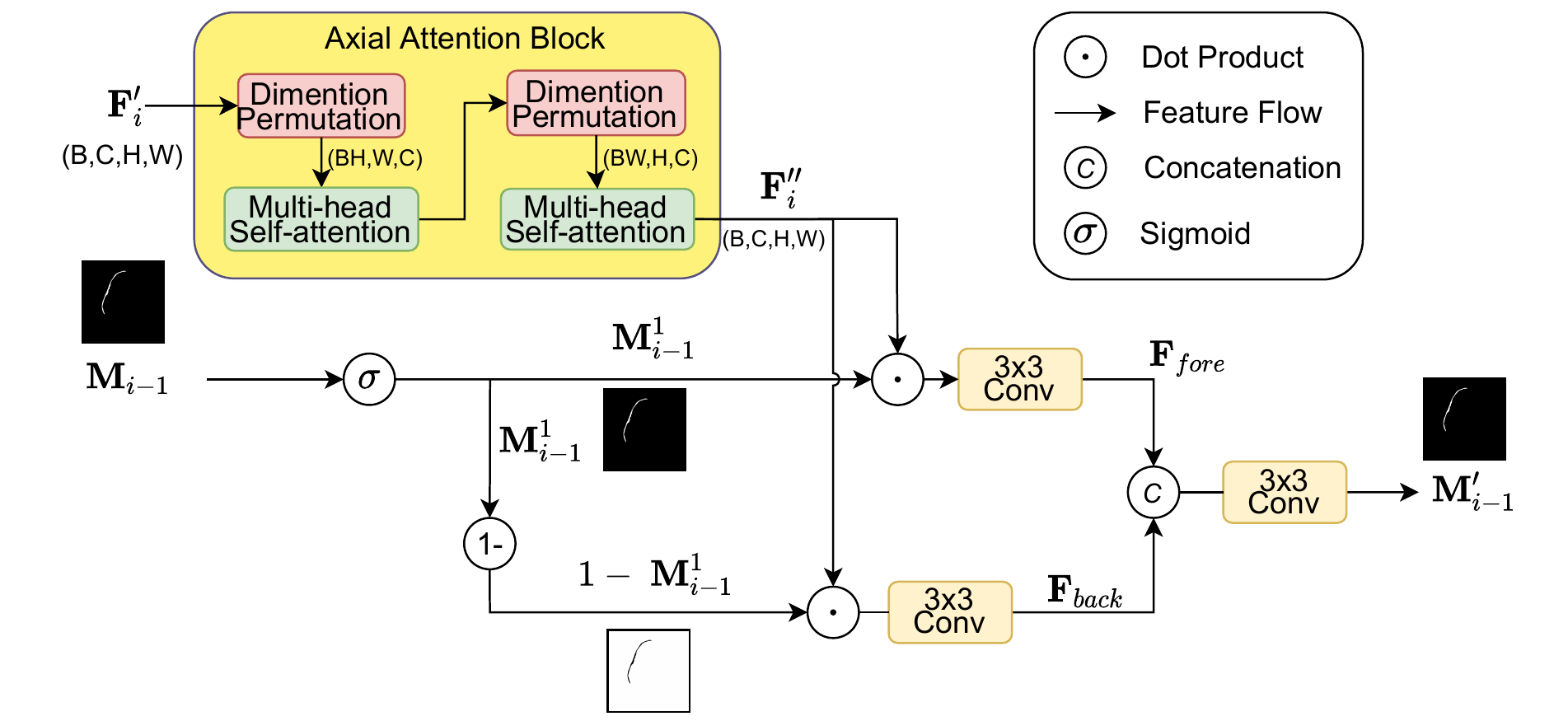}}
	\caption{Illustration of the Axial Context Relation Encoder (ACRE) block. The feature map $\mathbf{F}'_i$, $i=1,2,3$ indicates the output of MFI block in the Figure \ref{fig_1}. $\mathbf{F}'_i$ is first processed by Axial Attention Block, which contains two self-attention operations on different dimensions (H and W) through dimension permutation to get feature map $\mathbf{F}''_i$. The original mask $\textbf{M}_{i-1}$ is processed by the sigmoid function and then reproduced in two copies, one kept as is and one processed as $1-\textbf{M}_{i-1}$. After that, we multiply the $\mathbf{F}''_i$ with $\textbf{M}_{i-1}$ and $1-\textbf{M}_{i-1}$ respectively to get foreground $\mathbf{F}_{fore}$ and background $\mathbf{F}_{back}$ and then get the refined information $\textbf{M}'_{i-1}$ through concatenation of $\mathbf{F}_{fore}$ and $\mathbf{F}_{back}$ on channel level and convolution operation.}
	\label{fig_3}
\end{figure}

\subsection{The main structure}\label{AA}

Figure \ref{fig_1} shows our proposed CM-MLP framework, which consists of the 5-stage Encoder, the Partial Decoder, and the Parallel Branch. In the Parallel Branch, the MFI block can process all local information from the deeper layer of CNN (Convolutional Neural Network). The ACRE block can make the CM-MLP framework focus on exploring the boundary relationship between foreground and background. The MFI and ACRE block of the model is explained in detail below.

\subsection{Multi-scale Feature Interaction (MFI) block}\label{AB}

The application of MLP in visual tasks is mainly limited to image classification. By dividing the input image into non-overlapping patches and merging each patch in the spatial and channel dimensions to extract rich image information. Maxim's proposal\cite{b19} enables us to see that MLP has good performance on dense prediction tasks. It divides the original feature map into local and global branches aiming to extract information at different scales. Inspired by Maxim\cite{b19}, we propose a cascade Multi-scale MLP (Cascade MLP) in MFI block to encode the information and then fuse the information into a larger receptive field through Local MLP and Global MLP of multi-scale. 

Our proposed multi-scale Feature Interaction (MFI) block has two branches, in which the feature map $\textbf{F}$ is channel-wisely split into $\mathbf{F}_{up}, \mathbf{F}_{bottom}$. As shown in Figure \ref{fig_2}, three scales of Global MLP and Local MLP are connected in series to get the enriched features map gradually. 

In the Global MLP block (i.e., red block of Figure \ref{fig_2}), the input feature map (with size $(H, W, C)$) is grid into $(g\times g)$ non-overlapping patches of size $(H_g\times W_g)$ where $H = g\times H_g, W = g\times W_g$, to obtain the feature map $(g\times g$, $H_g\times W_g$,$C)$.

In the Local MLP block (i.e., orange block of Figure \ref{fig_2}), the feature map (with size $(H, W, C)$) is blocked into ($H_b\times W_b$) non-overlapping patches of size $(b\times b)$ where $H = b\times H_b, W = b\times W_b$, resulting in the feature map $(H_b\times W_b, b\times b, C)$. 

It is noted that the size of patches in Global MLP and Local MLP (i.e., $b$ and $g$) are not independent of each other. It is specified that $b\times g = W = H$ (for example: $H = W = 16$, $g_1=8$, $g_2=4$, $g_3=2$, and corresponding $b_1=2$, $b_2=4$, $b_3=8$). When the size of patches in Global MLP gradually decreases ($g$ gradually shrinks), the size of patches in Local MLP ($b$ gradually increases) increases. In other words, the distribution of the points in the FC input vector in Global MLP will be more sparse. In Local MLP, the number of points in the FC input vector in patches will be larger. Both of them will gradually expand the receptive field of the Cascade MLP. Therefore, The MFI block gradually fuses discrete local information and can get a gradually enriched features map. 




\subsection{Axial Context Relation Encoder (ACRE) block}\label{AC}

In order to make the MFI block focus on mining edge information instead of the body, we propose Axial Context Relation Encoder (ACRE) block inspired by\cite{b31}. The ACRE block focuses on the distinction between foreground and background boundaries so that the MFI block will extract more edge information. 


As shown in Figure \ref{fig_3}, each ACRE block has three main steps: firstly, the input feature map $\mathbf{F}'_i$ is sent into the axial attention block, which contains two self-attention operations on different dimensions (H and W) through dimension permutation, to obtain the feature map $\mathbf{F}''_i$. 

Secondly, $\mathbf{F}''_i$ will be masked by $\textbf{M}_{i-1}$ to obtain the foreground feature $\mathbf{F}_{fore}$ and the background feature $\mathbf{F}_{back}$: 
\begin{equation} \label{eq:mask1}
\begin{split}
\mathbf{F}_{back}=\phi_{back}(\mathbf{F}''_i\odot(1-\textbf{M}_{i-1})), 
\end{split}
\end{equation}
\begin{equation} \label{eq:mask2}
\begin{split}
\mathbf{F}_{fore} = \phi_{fore} (\mathbf{F}''_i \odot \textbf{M}_{i-1}),
\end{split}
\end{equation}
where $\phi_{back}(\cdot)$ and $\phi_{fore}(\cdot)$ represent the 3 × 3 convolutions and $\odot$ represents the dot product.

Finally, the output feature $\textbf{M}'_{i-1}$ of the ACRE block is obtained through the concatenation of $\mathbf{F}_{fore}$ and $\mathbf{F}_{back}$ on the channel dimension and $3\times 3$ convolution.

To ensure the effect of axial attention block on the output feature $\textbf{M}'_{i-1}$, we do not apply the offset in\cite{b31} to the contextual feature extraction of spatial locations. Second, the ACRE block can also be regarded as a supplement to the axial reverse attention in CaraNet\cite{b29}.



\subsection{Deep Supervision}\label{AD}

To better emphasize the segmentation task of each branch, we adopt a deep-supervised way to add the loss of each branch to the total loss. 

The loss function can be represented as $\ell=\ell_{IOU}+\ell_{BCE}$ by applying weighted intersection over union (IoU) and weighted binary cross entropy (BCE). we apply deep supervision for the four branch results ($\textbf{M}_0$, $\textbf{M}_1$, $\textbf{M}_2$, $\textbf{M}_3$). Before calculating the loss, we upsampled four branch results to the same size as ground truth $\textbf{G}$ as ($\textbf{M}^{up}_0$, $\textbf{M}^{up}_1$, $\textbf{M}^{up}_2$, $\textbf{M}^{up}_3$). Thus, the total loss can be represented as:

\begin{equation} \label{eq:loss}
\begin{split}
\ell_{total}=\sum\limits_{i=0}^3\ell(\textbf{G},\textbf{M}_i^{up})
\end{split}
\end{equation}

\section{EXPERIMENTS AND RESULTS} 

\subsection{Datasets and baselines}

Experiments are performed on CVC-ClinicDB dataset\cite{b32}, sub-Kvasir dataset\cite{b33}, and our in-house dataset. The CVC-ClinicDB dataset\cite{b32} is a polyps segmentation dataset, which contains 612 open-access images from 31 colonoscopy clips. The sub-Kvasir dataset\cite{b33} is a polyps segmentation dataset, which contains 1,000 images selected from a sub-class (polyp class) of the Kvasir dataset. Our in-house dataset is a large subdural hematoma segmentation dataset comprising 1049 images from 65 patients. All the three segmentation dataset is divided into training, validation and testing sets with the ratio of 7:1:2. We compare our proposed CM-MLP framework with four the-state-of-art medical image segmentation methods: U-Net\cite{b8}, U-Net++\cite{b4}, PraNet\cite{b28} and CaraNet\cite{b29}.

\subsection{Implementation details}
We implement our model in PyTorch. Affine transformation, horizontal flip and vertical flip are used for data augmentation. All the input images are uniformly resized to 512×512. LookAhead\cite{b34} optimization algorithm is used to optimize the parameters. The entire network is trained in an end-to-end way. Following the work CaraNet\cite{b29}, We employ three metrics (i.e., Mean Dice, Mean IoU and MAE) for quantitative evaluation and utilize MPA metrics to evaluate pixel-level accuracy.

\begin{table}[htbp]
	\caption{COMPARISON OF SEGMENTATION RESULTS ON THE SUB-KVASIR DATASET.}
	\label{kvasir_performance}
	\begin{center}
	    \resizebox{\linewidth}{!}{
		\begin{tabular}{c c|c|c|c|c}
			\midrule
			&$\textbf{Methods}$&$\textbf{Dice}$&$\textbf{mIoU}$&$\textbf{MAE}$&$\textbf{MPA}$ \\
			\midrule
			&U-Net\cite{b8}				&0.6321& 0.7704&0.0608&0.8949 \\
			&U-Net++\cite{b4} 			&0.8139& 0.7014&0.0372&0.8362 \\
			&PraNet\cite{b28}			&0.9454& 0.8970&0.0120&0.9508 \\
			&CaraNet\cite{b29}			&0.9482& 0.9027&0.0113&0.9595 \\
		&CM-MLP (Ours)	&$\mathbf{0.9676}$&$\mathbf{0.9373}$&$\mathbf{0.0087}$&$\mathbf{0.9658}$ \\
			\hline
		\end{tabular}
		}
		\label{tab1}
	\end{center}
\end{table}

\begin{table}[htbp]
	\caption{COMPARISON OF SEGMENTATION RESULTS ON THE CVC-CLINICDB DATASET.}
	\label{cvc_performance}
	\begin{center}
	    \resizebox{\linewidth}{!}{
		\begin{tabular}{c c|c|c|c|c}
			\midrule
   		&\textbf{Methods}&$\textbf{Dice}$&$\textbf{mIoU}$&$\textbf{MAE}$&$\textbf{MPA}$ \\
            			\midrule
			&U-Net\cite{b8}					&0.6469&0.4858&0.0544&0.6585 \\
			&U-Net++\cite{b4}  			&0.7290&0.5917&0.0357&0.8443 \\
			&PraNet\cite{b28}	 			&0.9420&0.8951&0.0071&0.9582 \\
			&CaraNet\cite{b29}				&0.9611&0.9256&0.0060&0.9665 \\
			&CM-MLP (Ours)	&$\mathbf{0.9696}$&$\mathbf{0.9412}$&$\mathbf{0.0048}$&$\mathbf{0.9758}$ \\
			\hline
		\end{tabular}
		}
		\label{tab1}
	\end{center}
\end{table}

\begin{table}[htbp]
	\caption{COMPARISON OF SEGMENTATION RESULTS ON OUR IN-HOUSE DATASET.}
	\label{inhouse_performance}
	\begin{center}
	    \resizebox{\linewidth}{!}{
		\begin{tabular}{c c|c|c|c|c}
			\midrule
			&$\textbf{Methods}$&$\textbf{Dice}$&$\textbf{mIoU}$&$\textbf{MAE}$&$\textbf{MPA}$ \\
            			\midrule

			&U-Net\cite{b8}				&0.7583& 0.6231&0.0035&0.6707 \\
			&U-Net++\cite{b4} 			&0.6997& 0.5595&0.0039&0.7175 \\
			&PraNet\cite{b28} 			&0.7949& 0.6649&0.0026&0.9315 \\
			&CaraNet\cite{b29}				&0.8155& 0.6940&0.0025&0.9249 \\
			&CM-MLP(Ours)	&$\mathbf{0.8254}$& $\mathbf{0.7087}$&$\mathbf{0.0024}$&$\mathbf{0.9378}$ \\
			
			\hline
			
		\end{tabular}
		}
		\label{tab1}
	\end{center}
\end{table}

\begin{table}[htbp]
	\caption{ABLATION STUDY OF OUR PROPOSED CM-MLP FRAMEWORK ON THE CVC-CLINICDB, SUB-KVASIR AND OUR IN-HOUSE DATASET.}
	\label{table-ablation}
	\begin{center}
	    \resizebox{\linewidth}{!}{
		\begin{tabular}{c c|c|c|c|c}
			\hline
			&\textbf{settings}&\textbf{Dice}&\textbf{mIoU}&\textbf{MAE}&\textbf{MPA} \\
			\hline
			
			\multirow{6}*{\rotatebox{90}{sub-Kvasir}}
			&CM-MLP &$\mathbf{0.9676}$& $\mathbf{0.9373}$&0.0087&$\mathbf{0.9658}$ \\
			&CM-MLP w/o MFI		&0.9634&0.9295&0.0090&0.9653\\
			&CM-MLP w/o Local	&0.9631&0.9290&$\mathbf{0.0086}$&0.9636 \\
			&CM-MLP w/o Global		&0.9663& 0.9350&0.0088&0.9647 \\
			&CM-MLP w/o ACRE		&0.9644&0.9313&0.0093&0.9653 \\
			\hline
			\multirow{6}*{\rotatebox{90}{\footnotesize{CVC-ClinicDB}}}
			&CM-MLP				&$\mathbf{0.9696}$& $\mathbf{0.9410}$&$\mathbf{0.0048}$&0.9758 \\
			&CM-MLP w/o MFI		&0.9683&0.9387&0.00054&$\mathbf{0.9759}$  \\
			&CM-MLP w/o Local		&0.9669&0.9361&0.0055&0.9753 \\
			&CM-MLP w/o Global	&0.9682& 0.9385&0.0053&0.9754 \\
			&CM-MLP w/o ACRE		&0.9687&0.9393&0.0055&0.9725 \\

			\hline
			\multirow{6}*{\rotatebox{90}{\footnotesize{In-house }}}
			&CM-MLP				&$\mathbf{0.8254}$& $\mathbf{0.7087}$&0.0024&$\mathbf{0.9378}$ \\
			&CM-MLP w/o MFI		&0.8148&0.6925&0.0024&0.9353 \\
			&CM-MLP w/o Local	&0.8203&0.6992&0.0023&0.9348 \\
			&CM-MLP w/o Global	&0.8164& 0.6964&$\mathbf{0.0022}$&$\mathbf{0.9378}$ \\
			&CM-MLP w/o ACRE &0.8072&0.6811&0.0024&0.9235 \\			
			\hline
			
		\end{tabular}
		}
		\label{tab2}
	\end{center}
\end{table}

\subsection{Performance Comparison}
We conduct the performance comparison of medical image segmentation task on Table \ref{kvasir_performance} , Table \ref{cvc_performance}, Table \ref{inhouse_performance}. In all three datasets, our proposed method outperforms the current state-of-the-art methods: U-Net\cite{b8}, U-Net++\cite{b4}, PraNet\cite{b28}, and CaraNet\cite{b29}.

In Table \ref{kvasir_performance} and Table \ref{cvc_performance}, our proposed CM-MLP framework outperforms the current state-of-the-art models in all metrics. In particular, the mIoU score of our proposed CM-MLP framework outperforms the previous method by 3.46\% and 1.56\% in the CVC-ClinicDB and sub-Kvasir datasets, respectively. The experiment results demonstrate that our proposed CM-MLP framework has a strong learning ability to segment polyps images with complex edges effectively. 

In Table \ref{inhouse_performance}, we report the results of current state-of-the-art methods with our proposed CM-MLP framework in a more challenging in-house dataset. Our proposed CM-MLP framework outperforms the current state-of-the-art models in all metrics. Particularly, the mIoU score of our proposed M-MLP framework outperforms the previous method by 1.47\%. Therefore, our proposed CM-MLP can perform better on more complex segmentation tasks.

\subsection{Component Analysis}\label{ablation}

We perform the following ablation studies on all used datasets to verify the effectiveness of each component in the CM-MLP framework: 
(1) \textbf{CM-MLP}: Our proposed CM-MLP framework (2) \textbf{CM-MLP w/o MFI}: Our proposed CM-MLP framework without MFI block; (3) \textbf{CM-MLP w/o Local}: Our proposed CM-MLP framework whose MFI block without local MLP; (4) \textbf{CM-MLP w/o Global}: Our proposed CM-MLP framework whose MFI block without global MLP; (5) \textbf{CM-MLP w/o ACRE}: Our proposed CM-MLP framework without ACRE block;

We have the following observations from the results in Table \ref{table-ablation}. First, our proposed approach \textbf{CM-MLP} outperforms the method \textbf{CM-MLP w/o MFI}, which demonstrates it is practical to use the MFI block to process all local information at the same time. Second, our approach \textbf{CM-MLP} also outperforms both \textbf{CM-MLP w/o Local} and \textbf{CM-MLP w/o Global}, which indicates that it is beneficial to use both Local MLP and Global MLP to catch the information from global to local. Third, our proposed approach \textbf{CM-MLP} is better than the method \textbf{CM-MLP w/o ACRE}, which demonstrates that it is helpful to focus on exploring
the boundary relationship between foreground and background. 

\begin{table}[htbp]
	\caption{COMPARISON RESULTS OF DIFFERENT CONNECTIONS OF MLPS ON THE CVC-CLINICDB, SUB-KVASIR AND OUR IN-HOUSE DATASET.}
	\label{connection}
	\begin{center}
	    \resizebox{\linewidth}{!}{
		\begin{tabular}{c c|c|c|c|c}
			\hline
			&\textbf{settings}&\textbf{Dice}&\textbf{mIoU}&\textbf{MAE}&\textbf{MPA} \\
			\hline
			\multirow{3}*{{sub-Kvasir}}
			&CM-MLP &$\mathbf{0.9676}$& $\mathbf{0.9373}$&0.0087&$0.9658$ \\

			&MFI-PP	&0.9674&0.9370&0.0088&$\mathbf{0.9725}$ \\
    		&MFI-CP	&0.9653& 0.9330&$\mathbf{0.0076}$&0.9640 \\
			\hline
			\multirow{3}*{{\footnotesize{CVC-ClinicDB}}}
			&CM-MLP				&$\mathbf{0.9696}$& $\mathbf{0.9410}$&$\mathbf{0.0048}$&$\mathbf{0.9758}$ \\		
			&MFI-PP	&$\mathbf{0.9696}$& $\mathbf{0.9410}$&0.0050&0.9719 \\	
			&MFI-CP		&0.9678& 0.9377&0.0055&0.9725 \\		
			\hline
			\multirow{3}*{{\footnotesize{In-house Dataset}}}
			&CM-MLP				&$\mathbf{0.8254}$& $\mathbf{0.7087}$&0.0024&$\textbf{0.9378}$ \\
			&MFI-PP		&0.8205& 0.7002&$\mathbf{0.0022}$&0.9291 \\
			&MFI-CP		&0.8200& 0.6986&0.0024&0.9357 \\
			\hline
		\end{tabular}
		}
		\label{tab2}
	\end{center}
\end{table}

\begin{table}[htbp]
	\caption{COMPARISON OF OUR PROPOSED MFI BLOCK WITH CFP BLOCK PROPOSED IN \cite{b29} ON THE CVC-CLINICDB, SUB-KVASIR AND OUR IN-HOUSE DATASET.}
	\label{CFP}
	\begin{center}
	    \resizebox{\linewidth}{!}{
		\begin{tabular}{c c|c|c|c|c}
			\hline
			&\textbf{settings}&\textbf{Dice}&\textbf{mIoU}&\textbf{MAE}&\textbf{MPA} \\
			\hline
			\multirow{2}*{{sub-Kvasir}}
			& CM-MLP  &$\mathbf{0.9676}$& $\mathbf{0.9373}$&0.0087&$\mathbf{0.9658}$ \\
			& CFP &0.9644& 0.9317&\textbf{0.0082}&0.9658 \\			
			\hline
			\multirow{2}*{{\footnotesize{CVC-ClinicDB}}}
			& CM-MLP &$\mathbf{0.9696}$& $\mathbf{0.9410}$&$\mathbf{0.0048}$&$\mathbf{0.9758}$ \\
			& CFP		&0.9673& 0.9368&0.0053&0.9370 \\
			\hline
			\multirow{2}*{{\footnotesize{In-house Dataset}}}
			& CM-MLP 				&$\mathbf{0.8254}$& $\mathbf{0.7087}$&$\textbf{0.0024}$&0.9378 \\
			& CFP		&0.8206& 0.7011&0.0026&$\textbf{0.9430}$ \\
			\hline
		\end{tabular}
		}
		\label{tab2}
	\end{center}
\end{table}

\begin{table}[htbp]
	\caption{COMPARISON OF OUR PROPOSED ACRE BLOCK WITH A-RA BLOCK PROPOSED IN \cite{b29} ON THE CVC-CLINICDB, SUB-KVASIR AND OUR IN-HOUSE DATASET.}
	\label{A-RA}
	\begin{center}
	    \resizebox{\linewidth}{!}{
		\begin{tabular}{c c|c|c|c|c}
			\hline
			&\textbf{settings}&\textbf{Dice}&\textbf{mIoU}&\textbf{MAE}&\textbf{MPA} \\
			\hline
			\multirow{2}*{{sub-Kvasir}}
			&CM-MLP &$\mathbf{0.9676}$& $\mathbf{0.9373}$& \textbf{0.0087}&$\mathbf{0.9658}$ \\
			&A-RA		&0.9568& 0.9181&0.0120&0.9623 \\
			\hline
			\multirow{2}*{{\footnotesize{CVC-ClinicDB}}}
			&CM-MLP				&$\mathbf{0.9696}$& $\mathbf{0.9410}$&$\mathbf{0.0048}$&$\mathbf{0.9758}$ \\
			& A-RA		&0.9561& 0.9169&0.0079&0.9623 \\
			\hline
			\multirow{2}*{{\footnotesize{In-house Dataset}}}
			&CM-MLP				&$\mathbf{0.8254}$& $\mathbf{0.7087}$&0.0024&\textbf{0.9378} \\
			& A-RA		&0.8226& 0.7030&\textbf{0.0023}&0.9309 \\
			\hline
		\end{tabular}
		}
		\label{tab2}
	\end{center}
\end{table}

\begin{figure*}[!t]
	\centering
	\subfloat{\includegraphics[scale=0.8]{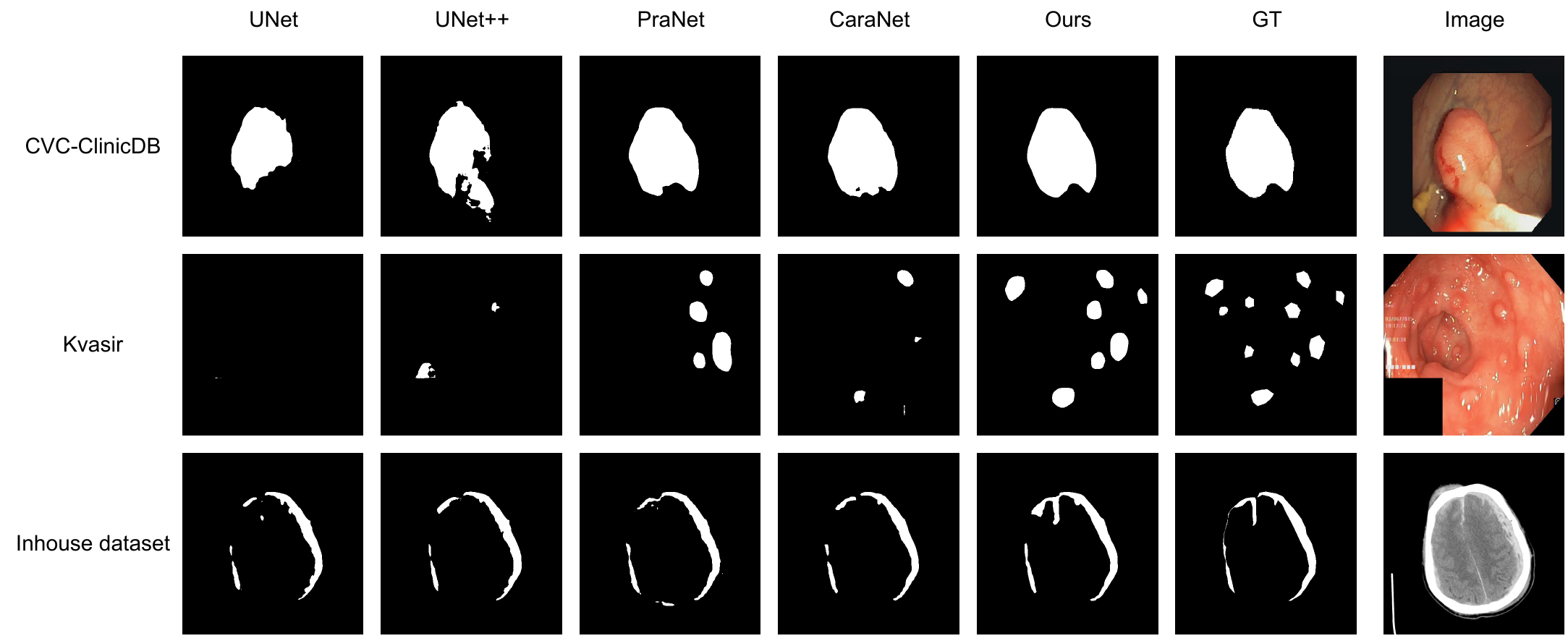}}
	\caption{Qualitative comparison of different methods in three datasets. From left to right: U-Net\cite{b8}, U-Net++\cite{b4}, PraNet\cite{b28},CaraNet\cite{b29},CM-MLP (Ours), Ground Truth image, and input segmentation image.}
	\label{fig_4}
\end{figure*}

In Table \ref{connection}, we report the results when comparing the different connections of Global MLP, Local MLP and Cascade MLP. Our proposed method \textbf{CM-MLP} use Cascade MLP in the MFI block in which Global MLP and Local MLP is connected in series. The method \textbf{MFI-PP} use parallel MLP between Local MLP and Global MLP proposed in Maxim \cite{b19} in Cascade MLP. For method \textbf{MFI-CP}, those three Cascade MLPs are paralleled with each other while keeping a series connection of Global MLP and Local MLP in each Cascade MLP. The comparison results demonstrate that the combined form of the Cascade MLP in the MFI block better captures the information when the segmentation tasks become difficult, which contains more complex edge of the segmented images.

In Table \ref{CFP}, we compare the effectiveness of our proposed MFI block with the \textbf{CFP} block proposed in \cite{b29}. The mIoU score of our proposed method is 0.5\%, 0.4\%, and 0.7\% higher than using \textbf{CFP} block in the sub-Kvasir dataset, CVC-ClinicDB dataset, and our in-house dataset respectively. The experiments demonstrate that the MFI block can better capture the local information using MLP than the \textbf{CFP} block proposed by \cite{b29}.

In Table \ref{A-RA}, it can be seen that after replacing the ACRE block with the axial reverse attention module (\textbf{A-RA}) proposed in \cite{b29}, the mIoU scores are significantly decreased by 1.9\% and 2.4\% in the sub-Kvasir dataset and CVC-ClinicDB dataset. This shows that the ACRE block is crucial to capturing edge information in sub-Kvasir and CVC-ClinicDB datasets. 


\subsection{Qualitative Analysis}
Fig \ref{fig_4} shows the results of the qualitative comparison, where our proposed CM-MLP framework provides better performance.

In the CVC-ClinicDB dataset, we observe that the previous state-of-the-art methods cannot capture the bump changes when the edge of the segmented images is more complex. In addition, the previous state-of-the-art methods and our proposed CM-MLP framework can successfully capture the body of the segmented images. These observations demonstrate that our proposed CM-MLP framework provides better performance when the edge of the segmented images is more complex and keeps the ability to capture the body of the segmented images simultaneously.

In the sub-Kvasir dataset, we observe that the current state-of-the-art methods cannot capture the body of the segmented images when the aggregation of multiple-segmented areas occurs. This demonstrates that our proposed CM-MLP framework can perform better when the aggregation of multiple-segmented areas appear in the medical image. In addition, the observation demonstrates that our proposed MLP-based MFI block can process all local information at the same time.

In our in-house dataset, we can see that the edges of the segmented images are more complex and the aggregation of multiple-segmented areas. Our proposed CM-MLP framework still captures the vital part of edges in the segmented image.

\section{CONCLUSION}

We propose a general MLP-based framework called CM-MLP for medical image segmentation. This framework can process all the local information of the image simultaneously. Therefore, the CM-MLP can cope with the complex edge information of a segmented area and the aggregation of multiple-segmented areas. While improving network performance minimizes the amount of computation and complexity brought by MLP. Extensive experiments demonstrate that our proposed CM-MLP consistently outperforms state-of-the-art methods on three challenging medical segmentation datasets. In the future, we will conduct our proposed CM-MLP on more datasets and explore the impact of data types from different modalities and sizes. 

\bibliographystyle{ieeetr} 
\bibliography{CM-MLP}

\end{document}